\documentclass[11pt,twocolumn,tighten]{aastex63}
\usepackage{hyperref}
\usepackage{color}
\usepackage{url}
\usepackage{natbib}
\usepackage{graphicx}
\usepackage{xfrac}
\usepackage{ulem}
\usepackage{CJK}
\usepackage{threeparttable}
\usepackage{multirow}
\usepackage{amsmath}
\begin{document}
\shorttitle{Galaxy-CGM co-evolution}
 \shortauthors{T.W. Lan}
\title{The Co-Evolution of Galaxies and the Cool Circumgalactic Medium  \\ Probed with the SDSS and DESI Legacy Imaging Surveys}
\begin{CJK*}{UTF8}{bsmi}
\author{Ting-Wen Lan 藍鼎文}
\affil{Kavli IPMU, the University of Tokyo (WPI), Kashiwa 277-8583, Japan}
\affil{Department of Astronomy and Astrophysics, UCO/Lick Observatory, University of California, \\ 1156 High Street, Santa Cruz, CA 95064, USA}
\begin{abstract}
We study the evolution of galaxies and the circumgalactic medium (CGM) through cosmic time by correlating $\sim50,000$ MgII absorbers, tracers of cool gas ($\sim10^{4}$~K), detected in the SDSS quasar spectra with galaxies detected in the DESI Legacy Imaging Surveys. By doing so, we extract the properties of galaxies associated with absorbers from redshift 0.4 to 1.3 with effectively $\sim 15,000$ pairs and explore the covering fraction of MgII absorbers as a function of galaxy type, stellar mass, impact parameter, and redshift. We find that the gas covering fraction increases with stellar mass of galaxies by $\sim M_{*}^{0.4}$. However, after we normalize the impact parameter by the virial radius of dark matter halos, the gas profiles around galaxies with masses ranging from $10^{9}$ to $10^{11} \, M_{\odot}$ become weakly dependent on stellar mass.
In addition, the gas distribution depends on galaxy type: the covering fraction within $0.3 \, r_{vir}$ around star-forming galaxies is 2-4 times higher than that around passive galaxies at all redshifts. We find that the covering fraction of strong absorbers ($W_{\lambda2796}>1 \rm \, \AA$) around both types of galaxies evolves significantly with redshift, similarly to the evolution of star-formation rate of galaxies, while such an evolution is not detected for weak absorbers ($W_{\lambda2796}<1 \rm \, \AA$). We quantify the HI mass traced by strong absorbers and find that the gas mass around galaxies evolves consistently with the star-formation rate of galaxies. This result suggests that the properties of galaxies and their CGM co-evolve through cosmic time.
Finally, we discuss the origins of strong absorbers around passive galaxies and argue that its redshift evolution may trace the star-formation activity of satellite galaxies.
 
\end{abstract}

\section{Introduction}

Gas flow processes, such as gas accretion and outflows, that regulate the amount of gas in and out of galaxies play an important role in driving the evolution of galaxies. These processes are expected to leave their signatures in gas around galaxies, the circumgalactic medium \citep[CGM,][for a review]{Tumlinson2017}. By studying the connection between galaxies and their CGM, one can therefore better understand how galaxies evolve through time. For several decades, absorption line spectroscopy has been used to probe the CGM via its absorption line features seen in spectra of background sources. MgII absorption lines $\lambda\lambda 2796, 2803 \rm \AA$ in particular have been used as one of the key tracers of cool gas ($\sim 10^{4} \rm \, K$) in the CGM because of their strength and visibility in optical spectra from redshift 0.4 to 2.5. 

Since the discovery of the first galaxy-MgII absorber pair \citep{Bergeron1986}, two approaches have been applied to investigate and characterize the relationships between MgII absorbers and their associated galaxies. The first approach involves performing targeted spectroscopic observations to identify and compile samples of galaxy-absorber pairs. By doing so, studies have probed the global distribution of gas around galaxies \citep[e.g.,][]{Steidel1994,Chen2010,Nielsen2013,Werk2013}, constrained the properties of gas associated with galactic outflows and inflows \citep[e.g.,][]{Glen2012, Bouche2012, Schroetter2016, Ho2017, Zabl2019} and revealed the small-scale structure of the CGM \citep[e.g.,][]{Rubin2018}. 
The second approach involves statistical analyses on large datasets from sky surveys. Novel relationships between galaxies and their CGM, such as the small-scale correlation between MgII absorption and galaxy properties \citep[e.g.,][]{Zibetti2007,Menard2011, Bordoloi2011,Zhu2013a,Lan2014, Lan2018}
and the large-scale gas distribution \citep[e.g.,][]{Zhu2014,Huang2016,Kauffmann2017}, have emerged with this approach.
While these studies have found interesting trends, most of the measurements are obtained at low redshifts ($z<1$) because it is expensive and challenging to perform similar measurements at higher redshifts. Therefore, how the galaxy-CGM connection evolves through cosmic time is still poorly known.

In this paper, we study the co-evolution of galaxies and the CGM from redshift 0.4 to 1.3 by cross-correlating MgII absorbers with galaxies detected in the DESI Legacy Imaging Surveys  \citep[hereafter the Legacy Surveys\footnote{\href{http://legacysurvey.org/}{legacysurvey.org}}][]{Dey2019}, which cover the footprint of the Sloan Digital Sky Survey \citep[SDSS, ][]{York2000} with 2 magnitudes deeper imaging data.  
The structure of 
the paper is as follows. Our data analysis is described in Section 2, and our results are presented in Section 3. The implications of our results are discussed in Section 4, and we summarize in Section 5. Throughout the paper we adopt a flat $\Lambda$CDM cosmology with $h=0.7$ and 
$\Omega_{\rm M}=0.3$.

\section{Data analysis}

\subsection{Datasets}

\textbf{Galaxy catalog:} To investigate the connection between galaxies and absorbers, we make use of a large galaxy catalog provided by the DESI Legacy Imaging Surveys \citep{Dey2019}. The Legacy Surveys consist of three public surveys, the Dark Energy Camera Legacy Survey, the Beijing-Arizona Sky Survey, and the Mayall z-band Legacy Survey. These surveys together cover about 14,000 deg$^{2}$ of the sky with g, r, and z bands.
The primary goal of the surveys is to provide the imaging dataset required for selecting galaxy candidates of the DESI spectroscopic survey \citep{Levi2013}. In this analysis, we make use of the DR8 catalog\footnote{\href{http://legacysurvey.org/dr8/description/}{legacysurvey.org/dr8/description/}}, in which galaxies are detected and their properties are characterized by the {\it Tractor} algorithm\footnote{\href{https://github.com/dstndstn/tractor}{github.com/dstndstn/tractor}} \citep{Lang2016} \citep[See also Section 8 in][]{Dey2019}. The catalog consists of about $9\times10^{8}$ galaxies. 

We further restrict our analysis within the footprint of the Legacy Surveys with at least three exposures in z-band and g-band and with the 5$\sigma$ limiting magnitude for extended sources in z-band being deeper than 22.5. The corresponding median limiting magnitudes of g and z bands for extended sources are about 24.2 and 23, which are about two magnitudes deeper than that of the SDSS.

\textbf{Metal absorber catalog:} For metal absorbers, we use the JHU-SDSS MgII absorber catalog\footnote{\href{https://www.guangtunbenzhu.com/jhu-sdss-metal-absorber-catalog}{www.guangtunbenzhu.com/jhu-sdss-metal-absorber-catalog}} provided by \citet{Zhu2013}. The MgII absorbers are detected from spectra of the SDSS I/II DR7 \citep{Schneiderdr7qso} and SDSS-III DR12 \citep{Paris2017} quasars via a fully automatic algorithm developed by the authors. We include absorber systems detected in both SDSS DR7 and DR12 quasar catalogs. In addition, we remove duplicated absorber entities detected in multiple observations of same quasars in both SDSS I/II and SDSS-III surveys. The sample contains $\sim$ 53,400 unique absorbers from redshift 0.4 to 2.5 with rest equivalent widths $W_{\lambda 2796}$ greater than 0.4 $\rm \AA$ in the selected footprint of the Legacy Surveys. There are $\sim$ 26,000 absorbers at redshifts 0.4 to 1.3, the main redshift region of our analysis.

\subsection{Methods}

\subsubsection{Mean number of galaxies associated with MgII absorbers}
To study the galaxy-CGM connection, we perform a cross-correlation analysis between MgII absorbers and the photometric galaxies detected in the Legacy Surveys. 
Here we adopt the method developed in \citet{Lan2014}.
We start from a set of absorbers (with similar absorption strength and redshift), count the number of photometric galaxies detected within a given physical aperture, and calculate the mean number of photometric galaxies per absorber with
\begin{equation}
    \langle N^{\rm Q(abs)}_{\rm gal} \rangle =  \frac{\sum_{j}^{N_{\rm abs}} N^{\rm Q(abs,j)}_{\rm gal}\times w}{N_{\rm abs}},
    \label{eq:n_gal}
\end{equation}
where $N_{\rm abs}$ is the total number of selected absorbers, $N^{\rm Q(abs,j)}_{\rm gal}$ is the number of galaxies around an absorber j, 
and $w$ is a correction factor for the galaxy number count close to the quasar line of sight ($<4"$) due to the blending between galaxies and quasars. The details of the functional form for $w$ are described in the Appendix. Here we assume that all the photometric galaxies are at the same redshifts of the absorbers. This mean galaxy number count around absorbers,$\langle N^{\rm Q(abs)}_{\rm gal} \rangle$, consists of two populations
\begin{equation}
    \langle N^{\rm Q(abs)}_{\rm gal} \rangle = \langle N^{\rm abs}_{\rm gal}\rangle + \langle N_{\rm bg} \rangle
    \label{eq:abs_gal}
\end{equation}
where $\langle N^{\rm abs}_{\rm gal}\rangle$ is from galaxies that are physically associated with absorbers and $\langle N_{\rm bg} \rangle$ is from background or foreground galaxies which are not associated with absorbers.

To estimate the contribution from unassociated galaxies, we perform the same procedure with a set of reference quasars which are selected to have 
SDSS i-band magnitude and redshift within 0.05 differences to the magnitude and redshift of the quasars with MgII absorbers. If fewer quasars satisfy these criteria, we increase the difference by 0.05 in both magnitude and redshift and search for reference quasars again. For each absorber, we select four reference quasars without any prior knowledge of the detection of absorbers along the line of sight. This selection ensures that the reference quasars are a representative sample that captures other galaxy contributions, including galaxies clustering with background quasars and galaxies associated with absorbers detected at different redshifts along the same line of sight,  which are not correlated with the presence of absorbers at the redshifts of  interest. With this reference set, we calculate the mean number of galaxies per reference quasar, $\langle N^{\rm Q(ref)}_{\rm gal} \rangle$, with
\begin{equation}
    \langle N^{\rm Q(ref)}_{\rm gal} \rangle =  \frac{\sum_{j}^{N_{\rm ref}} N^{\rm Q(ref,j)}_{\rm gal}\times w}{N_{\rm ref}},
    \label{eq:n_ref}
\end{equation}
and use it as an estimator of $\langle N_{\rm bg} \rangle$. 
By subtracting $\langle N^{\rm Q(ref)}_{\rm gal} \rangle$ from $\langle N^{\rm Q(abs)}_{\rm gal} \rangle$ in Eq.~\ref{eq:abs_gal}, we obtain the properties of galaxies associated with absorbers $\langle N^{\rm abs}_{\rm gal} \rangle$. 
Exploring $\langle N^{\rm abs}_{\rm gal} \rangle$ as a function of galaxy and absorber properties informs us on the connection between galaxies and their CGM. The uncertainties of the number counts are estimated by bootstrapping the absorber catalog 500 times.

\subsubsection{Covering fraction estimation}
In addition to investigating the galaxy population connected to absorbers,
it is crucial to characterize the incidence rate of absorbers around a galaxy population, e.g., covering fraction, $f_{c}$. It can be estimated as
\begin{equation}
    f_{c} = N_{\rm gal}^{[\rm abs]} / N_{\rm gal (\Delta z)}, 
    \label{eq:fc}
\end{equation}
where $N_{\rm gal}^{[\rm abs]}$ is the average number of galaxies connected to MgII absorbers per random sight-line within a certain aperture and $N_{\rm gal (\Delta z)}$ is the average number of galaxies at the same redshift range per random sight-line.

The numerator in Equation~\ref{eq:fc} can be estimated by 
\begin{equation}
    N_{\rm gal}^{[\rm abs]} = \langle N^{\rm abs}_{\rm gal} \rangle \times \frac{dN}{dz}(W_{\lambda2796})\times dz
\end{equation}
where $\frac{dN}{dz}(W_{\lambda2796})$ is the incidence rate of MgII absorbers adopted from  \citet{Zhu2013}. 

To estimate the denominator of Equation~\ref{eq:fc}, we use galaxy catalogs from two galaxy surveys: (1) the PRIsm MUlti-object Survey \citep[PRIMUS][]{Coil2011, Cool2013} and (2) UltraVISTA\footnote{\href{http://www.ultravista.org/}{www.ultravista.org/}}\citep{McCracken2012}. 
The PRIMUS survey is a galaxy redshift survey with magnitude limits similar to the magnitude limit of the Legacy Surveys.  
To obtain $N_{\rm gal (\Delta z)}$, we first count the number of PRIMUS galaxies with the selected redshift range and galaxy properties and divide it by survey area. We then scale it to the desired physical aperture size. When calculating the number of PRIMUS galaxies, instead of counting the raw number of galaxies in the catalog, we use the WEIGHT parameter of each galaxy provided by the catalog which accounts for the incompleteness due to target selection, collision of potential targets for spectroscopic observations, and redshift failures \citep[see][]{Moustakas2013}. In addition, we only use galaxies in COSMOS, XMM-SXDS, and XMM-CFHTLS fields 
because galaxies there within are selected with observed i-band brighter than 23 magnitude, a depth sufficient for our purpose. The total footprint of the three fields is $\sim3.2$ $\rm deg^{2}$. We use galaxy counts from PRIMUS to estimate $N_{\rm gal \, (\Delta z)}$ from redshift 0.4 to 1.

To estimate the covering fraction beyond redshift 1, we use the galaxy catalog\footnote{\href{https://www.strw.leidenuniv.nl/galaxyevolution/ULTRAVISTA/Ultravista/Catalog_Overview.html}{www.strw.leidenuniv.nl/galaxyevolution/ULTRAVISTA/Ultravista/}} from the UltraVISTA survey, which is 2-3 magnitudes deeper than the Legacy Surveys and covers $\sim1.6$ $\rm deg^{2}$ in the COSMOS field. We use the dataset of the first data release from \citet{Muzzin2013} which provides the photometric redshifts of galaxies derived from 30 bands and the physical properties of galaxies, such as stellar mass and absolute magnitudes via spectral energy distribution fitting. 
To calculate $N_{\rm gal (\Delta z)}$, we count the raw number of galaxies in the catalog given that with a limiting magnitude much deeper than the Legacy Surveys, the survey is expected to detect all the galaxies brighter than 22.5 magnitude in z-band.

\subsubsection{Stellar mass estimation}
In order to compare the galaxy-CGM connection as a function of redshift, we estimate 
the stellar mass of galaxies based on the observed $g-z$ galaxy color and z-band magnitude \citep[e.g.,][]{Mostek2012}. To do so, we use the UltraVISTA galaxy catalog from \citet{Muzzin2013, Muzzin2013b}. 
For each photometric galaxy around absorbers in the Legacy surveys, we first search UltraVISTA galaxies having photometric redshifts within 0.05 of the absorber redshifts. From the redshift-selected sample, we find the UltraVISTA galaxy that is closest to the photometric galaxy in $z$-band magnitude and $g-z$ color space and adopt the stellar mass of the UltraVISTA galaxy as the stellar mass of the galaxy. We remove galaxies around absorbers having the closest distance to its matched UltraVISTA galaxy in $z$-band and $g-z$ color space greater than 0.3 because those galaxies are expected to be background or foreground galaxies that are not at the redshift of absorbers. 
Additionally to the galaxies around absorbers, we apply the same process to the PRIMUS galaxies. This is for calibrating all the galaxies with the same fitting procedure. We note that the stellar mass measurements in the PRIMUS catalog \citep{Moustakas2013} are systematically 0.1 dex higher than the stellar mass measurements from the UltraVISTA catalog with $\sim0.2$ dex scatter.

Galaxies are further separated into star-forming and passive ones according to their $g-z$ colors. To obtain the color cuts, at redshift lower than 1, we use the stellar mass-star-formation rate (SFR) relation in \citet{Moustakas2013} and separate the PRIMUS galaxies into the two populations. 
We then identify the $g-z$ color cut that maximizes the separate of the two galaxy populations at each redshift bin. At redshift greater 1, we apply a similar procedure but use the UVISTA catalog and separate the two populations in $U-V$ and $V-J$ space following \citet{Muzzin2013b}. 
We note that at redshift lower 1, the color cuts from UltraVISTA are consistent with the cuts from PRIMUS.

\begin{figure}
\center
\includegraphics[width=0.5\textwidth]{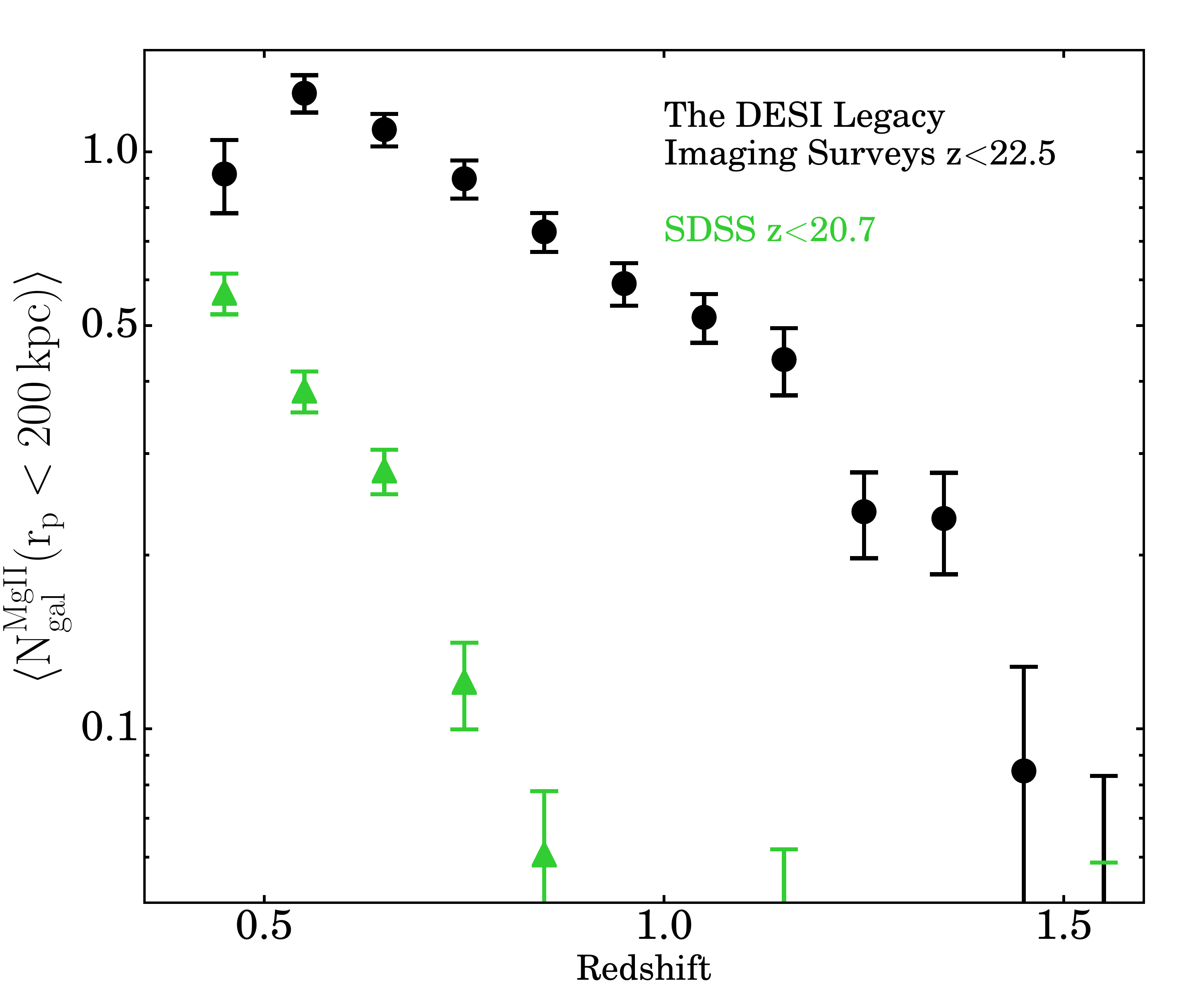}
\caption{Mean number of associated galaxies per absorber within 200 kpc as a function of redshift. The galaxy signal from the SDSS imaging dataset (z-band$<20.7$) and the signal from the Legacy Surveys (z-band$<22.5$) are shown by the green and black data points respectively.}
\label{fig:redshit_count}
\end{figure}

\begin{figure}
\center
\includegraphics[width=0.4\textwidth]{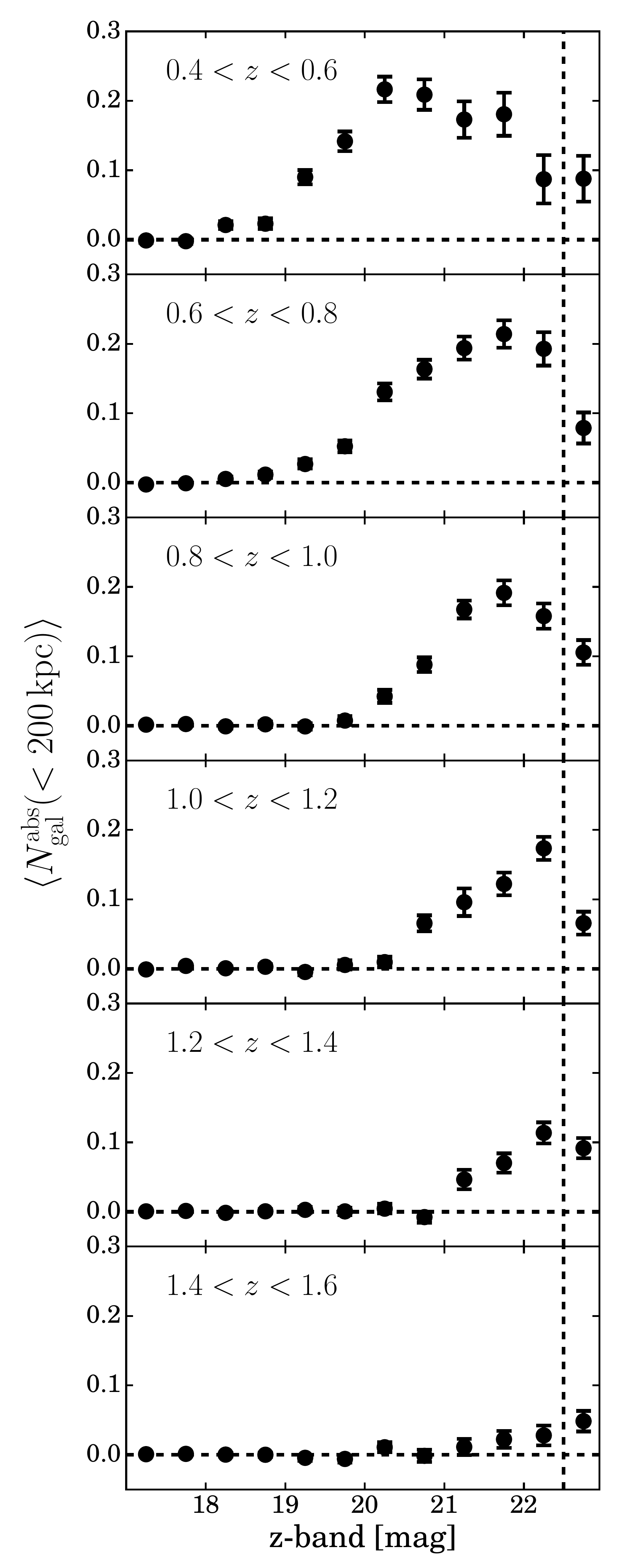}
\caption{Magnitude distribution of associated galaxies as a function of redshift. The 5-$\sigma$ limiting magnitude is 22.5 indicated by the vertical dashed lines.}
\label{fig:mag}
\end{figure}

\begin{figure*}
\center
\includegraphics[width=0.95\textwidth]{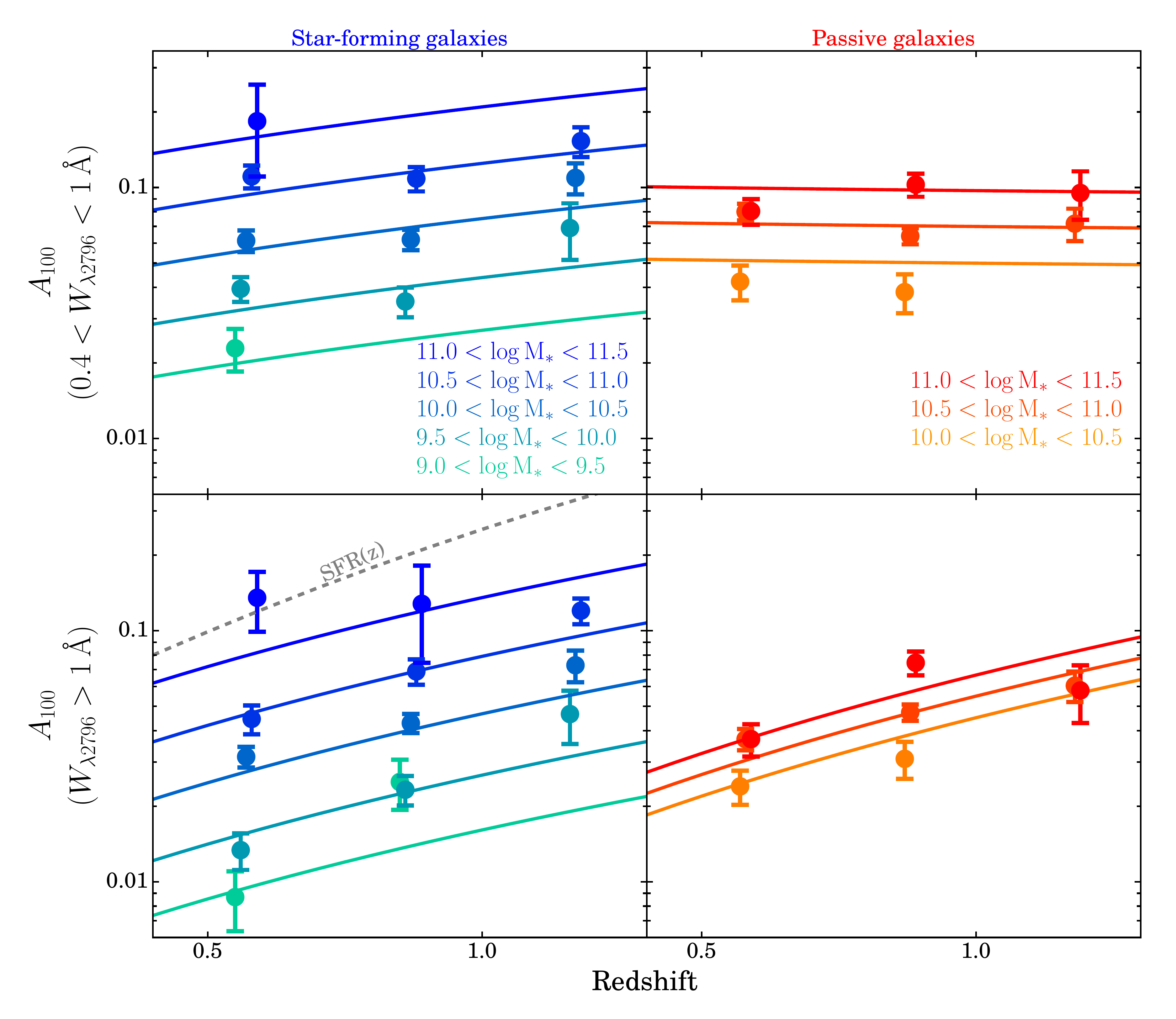}
\caption{Redshift evolution of the covering fraction of MgII absorbers. 
The data points show the best-fitting covering fraction at 100 kpc. 
The upper panel shows the results of weak absorbers ($0.4<W_{\lambda2796}<1 \rm \, \AA$) and the lower panel shows the results of strong absorbers ($W_{\lambda2796}>1 \rm \, \AA$). 
The left and right columns show the results for star-forming and passive galaxies respectively.
The color lines show the best-fit redshift and mass relation (Equation 6). The grey dashed line in lower-left panel shows the redshift evolution of star-formation rate of galaxies with a fixed stellar mass from \citet{Whitaker2012}. 
}
\label{fig:fc_large_para}
\end{figure*}

\section{Results}
\subsection{Observed properties of galaxies associated with MgII absorbers}
Being deeper than the SDSS, the Legacy Surveys are expected to detect fainter galaxies associated with absorbers. Figure~\ref{fig:redshit_count} shows the number of galaxies associated with absorbers within 200 kpc, $\langle N_{\rm gal}^{\rm abs}\rangle$, detected in the SDSS (green) and the Legacy Surveys (black) as a function of redshift. 
As can be seen, the signals obtained from the SDSS and the Legacy Surveys are significantly different; the SDSS detects galaxies up to redshift 0.8 while the Legacy Surveys detect galaxies up to redshift about 1.3. 
At $z<0.7$, the $\langle N_{\rm gal}^{\rm abs} \rangle$ values are consistent with 1, suggesting that the Legacy Surveys are deep enough to detect bulk of associated galaxies, if not all; this enables exploring the galaxy population connected to MgII absorbers.
The measurements are obtained by averaging over a few thousand MgII absorbers for each redshift bin. Together with the $\langle N_{\rm gal}^{\rm abs} \rangle$ values, we estimate the effective number of galaxy-absorber pairs within 200 kpc to be $\sim$15,000 from redshift 0.4 to 1.3, about two orders of magnitude larger than the sizes of previous samples obtained by targeted spectroscopic observations designed to identify individual galaxy-absorber pairs  \citep[e.g.,][]{Nielsen2013}.

Figure~\ref{fig:mag} shows the $z$-band magnitude distribution of galaxies associated with MgII absorbers ($W_{\lambda2796}>0.4 \, \rm \AA$) detected in the Legacy Surveys as a function of redshift. 
As expected, the brightness of associated galaxies becomes fainter at higher redshifts and the measurements become consistent with zero at redshifts higher than 1.5.

\begin{figure*}
\center
\includegraphics[width=1\textwidth]{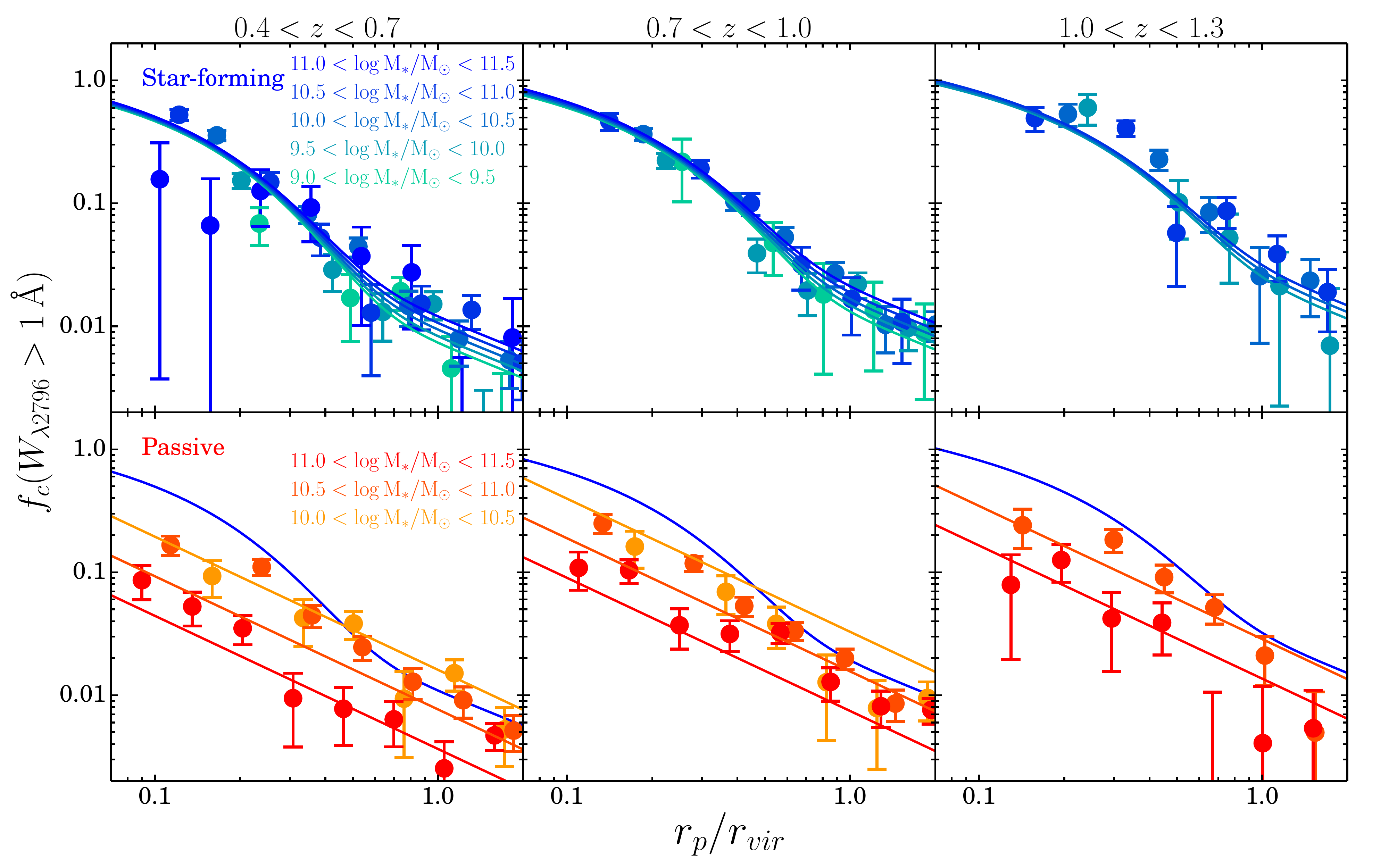}
\caption{Covering fraction of strong absorbers with impact parameters normalized by the virial radius of dark matter halos as a function of redshift. \emph{Top:} Gas around star-forming galaxies.  \emph{Bottom:} Gas around passive galaxies. The colors indicate the stellar mass of galaxies. The color lines show the best-fit functions for $f_{c}(<2 r_{vir})$. The best-fit profiles for star-forming galaxies with $10^{10.5} M_{\odot}$ are also shown in blue in the lower panels for comparison. }
\label{fig:fc_vir}
\end{figure*}

Below $z<0.6$, the magnitude distribution has an interesting shape; the mean number of galaxies peaks around 20 magnitude in $z$-band and decreases toward the faint end. Such a decline is an intrinsic property of associated galaxies. It is not due to the incompleteness of the imaging survey as the decline starts at around 20 magnitude which is 2.5 magnitudes brighter than the survey limiting magnitude indicated by the vertical dashed lines. 

We find that the shape of the magnitude distribution is inconsistent with the shape of the global magnitude distribution of galaxies at the same redshift with fainter galaxies dominating the number counts. We further explore the corresponding stellar mass distribution and find that MgII associated galaxies with stellar mass $>10^{10} \, M_{\odot}$ contribute to about $70\%$ of the total signal. 
This result illustrates that MgII absorbers preferentially trace galaxies with stellar mass $>10^{10}\, M_{\odot}$, consistent with clustering measurements  \citep[e.g.,][]{Bouche2006,Lundgren2009} showing galaxies associated with strong absorbers on average residing in dark matter halos with $\sim 10^{12}\, M_{\odot}$. It is also consistent with the inferred galaxy population, based from the stellar mass and metallicity relation \citep[e.g.,][]{Mannucci2009}, having sufficiently high metallicity to produce about solar metallicity of the circumgalactic MgII absorbers \citep{Lan2017}.

\subsection{Redshift evolution of gas distribution}
We now explore the distribution of cool gas traced by MgII absorbers around galaxies. To do so, we measure the covering fraction of MgII absorbers around galaxies  from 20 kpc to 600 kpc as a function of stellar mass, galaxy type, and redshift and fit the covering fraction with 
\begin{equation}
    f_{c} = A_{100} \times \bigg( \frac{r_{p}}{100 \rm \, kpc}\bigg)^{\gamma},
\end{equation}
where the $\gamma$ values are obtained from a global fitting for each galaxy type and rest equivalent width cut as listed in Table~\ref{table:slope}. In the Appendix, we show the measured covering fraction and the best-fit power law profiles.

Figure~\ref{fig:fc_large_para} shows the best-fit values of $A_{100}$ as a function of stellar mass and redshift from 0.4 to 1.3.  
We find that the gas distribution depends on the stellar mass; the trend is observed not only in galaxy types but also in absorption strength. In addition, we find that $A_{100}$ of strong absorbers for both star-forming and passive galaxies evolves significantly with redshift by increasing a factor of 3 from redshift 0.4 to 1.3 with a given stellar mass, while such a trend is not observed for weak absorbers. To quantify the redshift evolution and the stellar mass dependence, we characterize $A_{100}$ with 
\begin{equation}
    A_{100}= A\times\bigg(1+z\bigg)^{\alpha}\bigg(\frac{M_{*}}{10^{10} M_{\odot}}\bigg)^{\beta}.
    \label{eq:redshift}
\end{equation}
The best-fit relations are shown by the color lines in Figure~\ref{fig:fc_large_para} and the best-fit parameter values are listed in Table~\ref{table:a_100kpc}.

We find that the $A_{100}$ for star-forming galaxies scales with stellar mass of galaxies by $\sim M_{*}^{0.5}$, while the covering fraction around passive galaxies has a weaker dependence with $\sim M_{*}^{0.3}$.
This stellar mass dependence is consistent with the results of \citet{Chen2010} who reported that scaling the MgII gas distribution by the luminosity of galaxies with $\sim L_{B}^{0.35}$ reduced the scatters of gas distribution. 
It is also consistent with the results of \citet{Prochaska2014}, showing that the covering fraction of CII absorption lines increases with stellar mass of galaxies. 
We also note that the covering fraction of neutral hydrogen at local Universe obtained by \citet{Bordoloi2018} follows the same trend, showing that the covering fraction scales with stellar mass by $M_{*}^{0.3}$.

The parameter values for the redshift dependence, $\alpha$, are consistent with no evolution for weak absorbers around passive galaxies and with a weak evolution around star-forming galaxies $\sim(1+z)^{1}$, but they indicate significant evolution for strong absorbers with $\sim(1+z)^{2-2.5}$. These redshift trends are observed in both types of galaxies.
The redshift evolution indicates that in addition to stellar mass, there are more parameters that are connected to the covering fraction of strong absorbers. One possible parameter is the star-formation rate of galaxies which evolves similarly with the covering fraction of strong absorbers as shown by the grey dashed line in Figure~\ref{fig:fc_large_para}.
We will further explore such a possibility in Section~\ref{sec:mass}.

\subsection{Gas distribution in dark matter halos}
A fixed physical scale corresponds to different virial radius of dark matter halos. Therefore, in addition to characterize the covering fraction in physical space, we explore the gas distribution with respect to the size of dark matter halos. 
To do so, we obtain the halo mass based on the stellar mass-halo mass relation from \citet{Behroozi2013} and the virial radius based on the analytic formula from \citet{Bryan1998}. We normalize the impact parameters by the virial radius of dark matter halos.
Figure~\ref{fig:fc_vir} shows the covering fraction of strong absorbers ($W_{\lambda2796}>1\rm \, \AA$) around star-forming (top) and passive galaxies (bottom) as a function of redshift. The colors show the covering fraction measurements of galaxies with different stellar mass. We find that after the normalization, the covering fraction of star-forming galaxies with different masses now aligns with each other.
We also observe that the profiles of inner ($<0.5 r_{vir}$) and outer regions ($>0.5 r_{vir}$) have different slopes; this trend has also been seen in previous studies \citep[e.g.,][]{Lan2018}, showing that the covering fraction at small scales around star-forming galaxies is systematically higher than the extrapolation from the best-fit profile obtained at larger scales. 

To characterize the inner and outer regions of the covering fraction of strong absorbers around star-forming galaxies as a function of redshift, we use a global function with a combination of an exponential and a power law profiles:
\begin{equation}
    f_c = f_{c}^{\rm exp}+f_{c}^{\rm power}, 
\end{equation}   
where  
\begin{equation}
    f_{c}^{\rm exp}={\rm exp}\bigg({\frac{-r_{p}/r_{vir}}{c(1+z)^{d}}}\bigg) \, \, \&
\end{equation}     
\begin{equation}
    f_{c}^{\rm power}=A\times\bigg(1+z\bigg)^{\alpha}\bigg(\frac{M_{*}}{10^{10} M_{\odot}}\bigg)^{\beta}\bigg(\frac{r_{p}}{r_{vir}}\bigg)^{\gamma}.
\end{equation}
The best-fit profiles are shown with the color dashed lines in the upper panel of Figure~\ref{fig:fc_vir}. 
We apply the same function form to fit the gas profiles of strong absorbers around passive galaxies and find that the power law $f_{c}^{\rm power}$ alone is sufficient to describe the profiles around passive galaxies as shown in the lower panel of Figure~\ref{fig:fc_vir}. Similarly, we also find that the power law $f_{c}^{\rm power}$ alone is sufficient to describe the gas profiles of weak absorbers around both types of galaxies. The best-fit values are listed in Table~\ref{table:a_rv}. 

For strong absorbers around star-forming galaxies, we find that both $f_{c}^{\rm exp}$ and $f_{c}^{\rm power}$ evolve with redshift. We find that the covering fraction of strong absorbers around passive galaxies evolves with redshift by $(1+z)^{4\pm0.4}$, similar to evolution of $f_{c}^{\rm power}$ for star-forming galaxies.   
For weak absorbers, the covering fraction does not show strong evolution with redshift similarly to the measurements in physical space.

In contrast to the redshift evolution, we find that after the normalization, the covering fraction becomes weakly dependent on the stellar mass around star-forming galaxies for both weak and strong absorbers. This trend was also reported in \citet{Churchill2013,Chuchill2013b}. 
However, the covering fraction correlates with stellar mass negatively around passive galaxies ($M^{-0.6}$ for strong absorbers and $M^{-0.3}$ for weak absorbers) as shown in the lower panel of Figure~\ref{fig:fc_vir}. We note that this decreasing trend is driven by the most massive bin $11<log(M_{*}/M_{\odot})<11.5$. If the bin is excluded from the fitting procedure, 
the stellar mass dependence of the covering fraction becomes weaker with $M_{*}^{-0.3}$ for strong absorbers and  consistent with no correlation for weak absorbers. 

The power law components for strong absorbers around star-forming galaxies and passive galaxies are consistent with each other in terms of the radial profile and the redshift evolution. The consistency suggests that there is a common mechanism, which is not sensitive to the current star-formation activity of the central galaxies, giving rise to the absorption. On top of this gas absorption, another mechanism connecting to the SFR of the central galaxies produces extra gas absorption in the inner regions of the halos of star-forming galaxies, which is captured by $f_{c}^{exp}$. We note that within about $0.3 r_{vir}$, the covering fraction of strong absorbers around star-forming galaxies is about 2-4 times higher than the covering fraction around passive galaxies. The difference between the gas covering fraction around star-forming and passive galaxies can be seen in the lower panel of Figure~\ref{fig:fc_vir}, where the best-fit gas profile around star-forming galaxies is shown with the blue lines. 
This demonstrates that the dichotomy of galaxy types is reflected in cool gas traced by strong absorbers in the CGM at all redshifts, consistent with previous findings at low redshifts.

Previous studies \citep[e.g.,][]{Bordoloi2011, Zhu2013, Lan2014, Lan2018, Martin2019} have shown that the excess absorption around star-forming galaxies tends to be found along the minor axis of galaxies, suggesting that the gas is likely associated with galactic outflows. In this case, 
the gas profiles around star-forming galaxies contain crucial information for constraining the initial conditions of gas outflows such as velocity and ejected mass \citep[e.g.,][]{Lan2019}. 
The redshift evolution of $f_{c}^{exp}$ further suggests that at high redshifts, the gas associated with galactic outflows may travel farther away from the galaxies. 
In the following, we will estimate the amount of gas mass around galaxies and how the gas mass in the halos evolves with redshift.
\begin{figure*}
\center
\includegraphics[width=1.\textwidth]{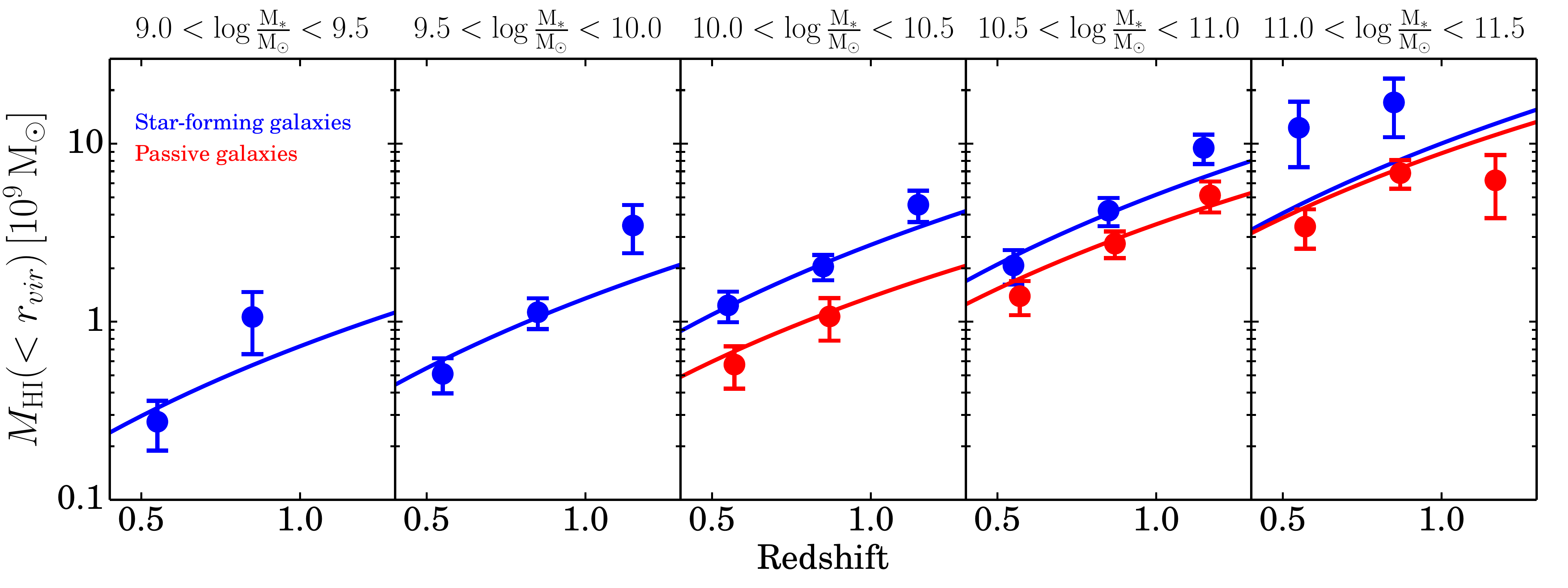}
\caption{Evolution of neutral hydrogen mass traced by strong absorbers ($W_{\lambda2796}>1 \rm \, \AA$) within the virial radius as a function of redshift and stellar mass for star-forming (blue) and passive (red) galaxies. The color lines show the best-fit relations of Equation 13.}
\label{fig:gas_fraction}
\end{figure*}


\subsection{Evolution of the HI mass and its connection to the SFR of galaxies}
\label{sec:mass}
We now estimate the neutral hydrogen mass traced by MgII absorbers in the CGM within the virial radius with
\begin{equation}
    M_{\rm HI}(<r_{vir})\sim 2\pi \, m_{\rm H} \int_{10 \, \rm kpc}^{r_{vir}}
    \hat{N}_{\rm HI}\, f_{c}(r_{p})\, r_{p}\, dr_{p}.
\end{equation}
where $N_{\rm HI}$ is taken from the empirical relationship between $W_{\lambda 2796}$ and $N_{\rm HI}$ in \citet{Lan2017}:
\begin{equation}
    \hat{N}_{\rm HI} = 10^{18.96}\, \bigg(\frac{W_{\lambda 2796}}{1 \rm \AA}\bigg)^{1.69} \bigg(1+z\bigg)^{1.88} \rm cm^{-2}.
    \label{eq:NHI}
\end{equation}
We note that while the scattering of $N_{\rm HI}$ for individual absorbers is about 0.5-1 dex with a given $W_{\lambda 2796}$ (see Figure 1 in \citet{Lan2017}), here we only consider the median amount of $N_{\rm HI}$ traced by MgII absorbers and the corresponding redshift evolution. The redshift evolution of $N_{\rm HI}$ in MgII absorbers is significantly detected in \citet{matejek2013} and \citet{Lan2017}. 
The statistical uncertainty of Eq~\ref{eq:NHI} is included in the estimation of $M_{\rm HI}$. 

To estimate $\hat{N}_{\rm HI}$, we use the median $W_{\lambda 2796}$ values from the absorber sample of \citet{Zhu2013} which are $\sim0.65 \rm \, \AA$ for weak absorbers and $\sim1.5 \rm \, \AA$ for strong absorbers. Given that the gas distribution traced by strong absorbers shows a significant evolution with redshift, in the following we focus on $M_{\rm HI}$ traced by strong absorbers.

Figure~\ref{fig:gas_fraction} shows the amount of neutral hydrogen traced by strong absorbers ($W_{\lambda2796}>1 \rm \AA$) in the cool CGM as a function of stellar mass and redshift. 
The $M_{\rm HI}$ increases with stellar mass around both star-forming galaxies and passive galaxies, while with a fixed stellar mass, star-forming galaxies have more neutral gas than passive ones. Such a difference is due to the excess absorption in the inner regions of cool gas traced by strong absorbers. 
Moreover, with a fixed stellar mass, the neutral hydrogen mass increases with redshift. We parameterize the redshift and stellar mass dependence with 
\begin{equation}
    \frac{M_{\rm HI}}{M_{\odot}} = A\times10^{9}\, \bigg(1+z\bigg)^{\alpha} \bigg(\frac{M*}{10^{10} M_{\odot}}\bigg)^{\beta}
\end{equation}
where the best fit values of the parameters are listed in the bottom part of Table~\ref{table:a_100kpc}.

\begin{figure*}
\center
\includegraphics[width=1\textwidth]{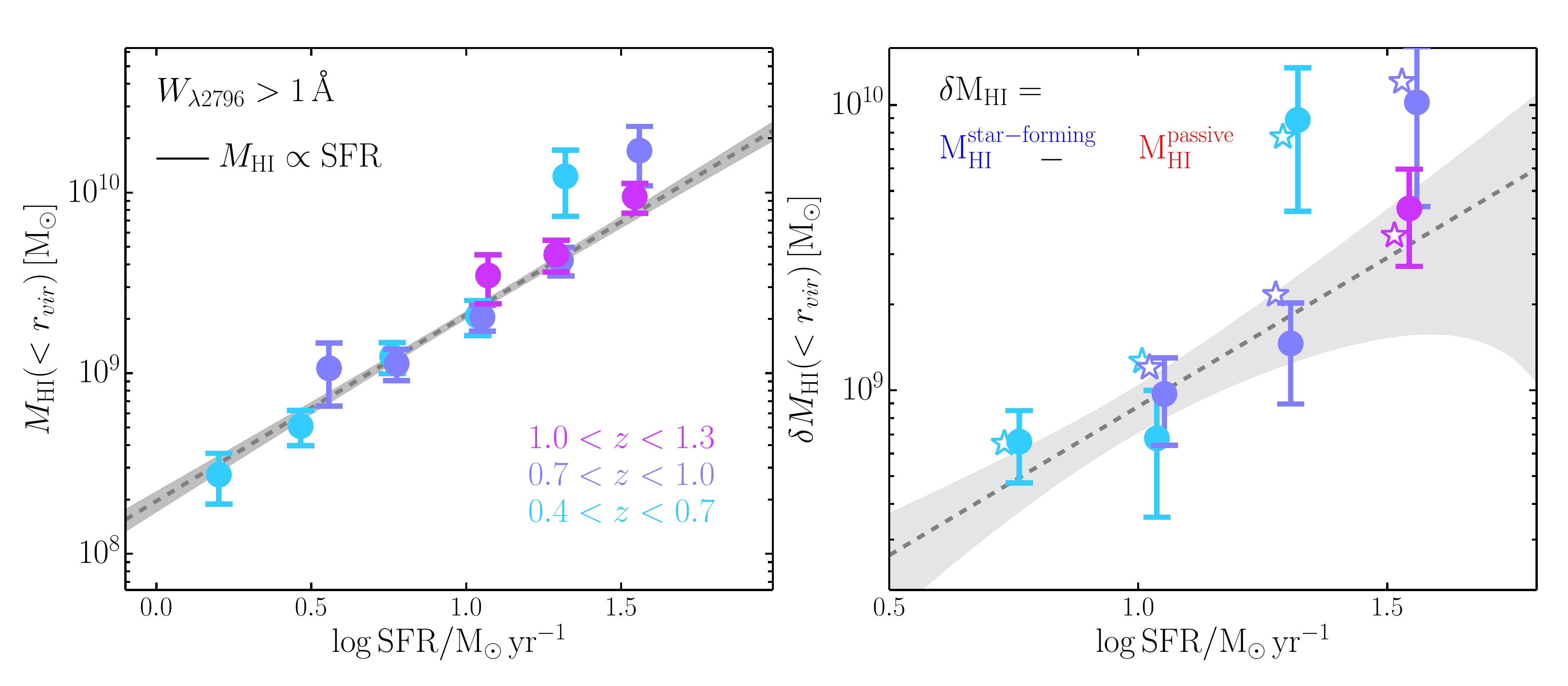}
\caption{Neutral hydrogen mass, $M_{\rm HI}$, traced by strong absorbers as a function of star-formation rate. \emph{Left:} Total neutral hydrogen mass within $r_{vir}$ around star-forming galaxies. \emph{Right:} Excess neutral hydrogen mass, $\delta M_{\rm HI}$,  connecting to the star-formation rate of the central galaxies. The dashed lines show the best-fit relations. The open stars in the right panel show the mass estimated based on the exponential profile $f_{c}^{exp}$ listed in Equation 9.}
\label{fig:SFR_mass}
\end{figure*}

We find that the redshift evolution of the amount of neutral hydrogen traced by strong absorbers around star-forming galaxies, $\sim(1+z)^{3}$, is consistent with the redshift evolution of star-formation rate of galaxies \citep[e.g.,][]{Whitaker2012,Lee2015,Scoville2017}.
To demonstrate that, we explore the correlation between $M_{\rm HI}$ and SFR by adopting 
the average SFR of galaxies for each stellar mass and redshift bin based on the empirical relation of the SFR-stellar mass as a function of redshift described in \citet{Whitaker2012}: 
\begin{eqnarray}
{\rm log\bigg(\frac{SFR}{M_{\odot}\, yr^{-1}}\bigg)} &=& (0.7-0.13z)\times({\rm log} \, M_{*}/M_{\odot}-10.5)
\nonumber\\
&&+\, (0.38+1.14z-0.19z^{2}),
\end{eqnarray}
where $z$ is the redshift of galaxies.

The left panel of Figure~\ref{fig:SFR_mass} shows $M_{\rm HI}(<r_{vir})$ traced by strong absorbers around star-forming galaxies as a function of SFR. As can be seen, $M_{\rm HI}(<r_{vir})$ values at different redshifts now align on a sequence, which can be described by a power law,
\begin{equation}
   M_{\rm HI}(<r_{vir})=(2.0\pm0.1)\times10^{9}\bigg(\frac{\rm SFR}{10 \,M_{\odot} \rm yr^{-1}}\bigg)^{1.0\pm0.1} M_{\odot}, 
    \label{eq:SFR_A}
\end{equation}
as shown by the grey line.     
This result shows that the redshift evolution of neutral hydrogen mass traced by strong absorbers and the SFR of galaxies are consistent with each other.

While the correlation between $M_{\rm HI}$ and SFR may suggest a connection between the cool gas in the CGM and the star-formation activity of the central galaxies, we note that the neutral hydrogen mass traced by strong absorbers around passive galaxies also evolves consistently with the evolution of SFR. 
Given that the central galaxies are red and passive with no expected recent star-formation activity, the evolution of the CGM may not directly link to the evolution of star-formation activity of the central galaxies. Instead, it could link to the star-formation activity of satellite galaxies which are expected to evolve with redshift similarly to star-forming galaxies in the field \citep[e.g. See Fig. 3 in ][]{Popesso2012}. While the origins of the cool gas around passive galaxies are still under debate, the cool gas associated with satellite galaxies, if not destroyed, is expected to be one of the sources contributing to the gas halos. In Section 4.2, we will explore this scenario and show that the cool gas around blue satellites could explain a large fraction of gas seen around massive passive galaxies. 
In other words, the redshift evolution of gas traced by strong absorbers around passive galaxies may reflect the evolution of star-formation activity in dark matter halos. 

To isolate the amount of gas associated only with the star-formation rate of the central galaxies, we calculate the excess amount of gas mass, $\delta M_{\rm HI}$, around star-forming galaxies by subtracting $M_{\rm HI}$ around passive galaxies from $M_{\rm HI}$ around star-forming galaxies. This is done only for the mass ranges that have the gas mass measurements for both types of galaxies. The right panel of Figure~\ref{fig:SFR_mass} shows the result, indicating that the excess amount of neutral gas also correlates with SFR and can be described as 
\begin{equation}
   \delta M_{\rm HI}(<r_{vir})=(0.9\pm0.2)\times10^{9}\bigg(\frac{\rm SFR}{10 \,M_{\odot} \rm yr^{-1}}\bigg)^{1.0\pm0.3} M_{\odot}.  
    \label{eq:SFR_mass_central}
\end{equation}
This result demonstrates that {\it $\sim 50\%$ of cool neutral gas mass traced by strong absorbers around star-forming galaxies directly links to the star-formation activity of the central galaxies, an evidence for the co-evolution of galaxies and their CGM through cosmic time.}

We note that the excess amount of gas, $\delta M_{\rm HI}(<r_{vir})$, is contributed by the inner region of cool gas traced by strong absorbers around star-forming galaxies captured by $f_{c}^{\rm exp}$ (Eq. 9). 
We can also estimate the mass based on the best-fit global gas profile $f_{c}^{\rm exp}$ of strong absorbers. The open stars in the right panel of Figure~\ref{fig:SFR_mass} are the estimations from the best-fit $f_{c}^{\rm exp}$, which are consistent with the measured ones.

As we pointed out in Section 3.3, the exponential component may trace the gas absorption associated with galactic outflows based on the azimuthal angle dependence of gas absorption as found in previous studies \citep[e.g.,][]{Bordoloi2011, Lan2018}.
In addition, the ejected mass and velocity of gas outflows increase with the SFR of galaxies \citep[e.g.,][]{Martin2012,Rubin2014}. Taken together, we argue that 
the correlation between $\delta M_{\rm HI}$ and SFR in the right panel of Figure~\ref{fig:SFR_mass} is consistent with the outflow scenario - galaxies with high SFR eject more gas materials with high velocities which propagate farther away from the galaxies. In other words, the gas profiles contain crucial information for constraining models of feedback mechanisms.

\section{Discussion}

\subsection{Comparisons with previous studies}
By cross-correlating MgII absorbers with galaxies detected in the Legacy Surveys, we have probed the galaxy population associated with MgII absorbers. In addition, we have estimated the covering fraction and showed that the covering fraction and the neutral hydrogen mass of strong absorbers evolve significantly with redshift and this evolution follows the evolution of star-formation rate of galaxies, suggesting a connection between the distribution of MgII absorbers and star-formation. These results are in agreement with results from other probes:

\textbf{(1) Redshift evolultion of the incidence rate of strong absorbers:} the incidence rate of strong MgII absorbers have been known to increase toward high redshift, a trend similar to the evolution of cosmic star-formation density of the Universe \citep[e.g.,][]{Prochter2006, Zhu2013, Matejek2012}. This similarity may suggest a physical connection between the two entities. In addition, \citet{Menard2011} showed that the surface brightness of [OII] emission associated with MgII absorbers observed in the fiber of SDSS quasars correlates with the absorption strength and redshift. Taking the evolution of dN/dz and the [OII] emission together, \citet{Menard2011} first showed that the cosmic star-formation density can be reproduced via absorption line statistics. \citet{Matejek2012} and \citet{Chen2017} further showed that this consistency between the cosmic star-formation density and the one inferred from MgII absorbers extends to redshift 5. 
    
To understand the dN/dz of strong absorbers, \citet{Chen2017} calculated the expected incidence rate by integrating dark matter halo mass function together with assumed gas distributions in the halos. They found that the model with the gas distribution scaling with halo virial radius and with the integrated minimum halo mass based on star-formation rate can best reproduce the incidence rate of MgII absorbers. This result is consistent with our findings.

\textbf{(2) Physical properties of MgII absorbers:} The physical properties of MgII absorbers also provide information about the associated galaxies. \citet{Lan2017} measured the metallicity and the gas volume density of MgII absorbers and showed that the average metallicity of MgII absorbers is approximately solar metallicity at redshift $z<1$. Based on the galaxy mass and metallicity relation \citep[e.g., Fig. 5 in ][]{Mannucci2009}, this metallicity corresponds to galaxy stellar mass $M_{*}>10^{10} \, M_{\odot}$, consistent with the observed magnitude distribution of the associated galaxies as shown in Figure~\ref{fig:mag}. Furthermore, the hydrogen volume density of MgII absorbers is $\sim0.3\, \rm  cm^{-3}$ reported in \citet{Lan2017}. If we assume that the cool gas is in pressure equilibrium with the surrounding hot gas, it would require hot gas around $>10^{12} \, \rm M_{\odot}$ halos ($M_{*}\sim10^{10.5} M_{\odot}$ galaxies) to provide sufficient pressure support \citep[e.g.,][]{Mo1996, Maller2004,Lan2019} for the observed gas volume density of MgII absorbers. 

We note that our results are not consistent with the results of \citet{Chen2012}, showing that the average absorption of MgII around L* galaxies does not evolve significantly from redshift 0.3 to 2.2. While new additional data is required to resolve the conflict, we propose that this discrepancy could be due to the following factors: 
\begin{itemize}
    \item In our results, we find that only the covering fraction of strong absorbers show strong redshift evolution, while the covering fraction of weak absorbers does not. \citet{Chen2012} adopted the measurements of \citet{Bordoloi2011} and \citet{Steidel2010}, which were obtained by stacking all the spectra. Therefore, the average absorption strengths are contributed by both weak and strong absorbers. This will dilute the signal of redshift evolution.
    \item In addition, the redshift 0.25 measurements from \citet{Chen2012} were obtained with background quasars, while the redshift 0.7 \citep{Bordoloi2011} and redshift 2.2 \citep{Steidel2010} measurements were obtained with background galaxies. Because of the physical sizes of background objects, these measurements may not probe the same physical structure of the CGM. We also note that the spectral resolution of zCOSMOS is too low to resolve the doublet of MgII lines, which could also dilute the absorption strength signal. 
\end{itemize}
We emphasize that all the factors described above may dilute the redshift evolution as shown in this work and may result the inconsistency between the results of our work and \citet{Chen2012}.

\subsection{Origin(s) of MgII absorbers}
The distribution of MgII absorbers around galaxies and its redshift evolution add new information to disentangle the mechanisms that give rise to the absorption line systems. Despite that there is no distinct feature in the rest-equivalent width distribution dN/dW of MgII absorbers, the evolution of weak ($W_{\lambda2796}<1 \rm \, \AA$) and strong absorbers ($W_{\lambda2796}>1 \rm \, \AA$) and their connection to galaxies are significantly different. As shown in the top panel of Figure~\ref{fig:fc_large_para}, the covering fraction of weak absorbers around both star-forming and passive galaxies depends on stellar mass, suggesting that a physical mechanism purely connected to the mass of galaxies regulates the distribution of weak absorbers. 

In contrast to weak absorbers, the covering fraction of strong absorbers evolves with redshift. It would require additional mechanisms to explain such an evolution. While in Figure~\ref{fig:SFR_mass}, we show that the redshift evolution of the covering fraction of strong absorbers follows the redshift evolution of star-formation rate of galaxies, the fact that the covering fraction around passive galaxies evolves similarly to the covering fraction around star-forming galaxies suggests that the covering fraction of strong absorbers may not trace the star-formation activity of central galaxies. Instead it may trace the global star-formation rate in the vicinity of galaxies, including the contribution of all the satellite galaxies (one-halo) and neighboring halos (two-halo). 

To test this scenario, we perform a simple estimation for the gas cross-section around passive galaxies and the expected contribution from star-forming satellite galaxies. 
We calculate the gas cross-section around passive galaxies with
\begin{equation}
    \sigma_{\rm passive \, galaxies}(<r_{vir})\sim 2\pi \int^{r_{vir}} f_{c}(r_{p})\, r_{p}\, dr_{p}.
\end{equation}
If we consider the covering fraction of strong absorbers observed around passive galaxies with $M_{*}\sim10^{11} M_{\odot}$ at redshift 0.6, the corresponding $r_{vir}$ is about 400 kpc
and the $\sigma_{\rm passive \, galaxies}$ is about $6000 \, \rm kpc^{2}$.

We now estimate the gas cross-section from the CGM of satellite galaxies. 
We use the measured stellar mass function of star-forming satellite galaxies from \citet{Lan2016} and the best-fit exponential $f^{exp}_{c}$ to estimate the total gas cross-section from satellite galaxies. We find that the total cross-section of galaxies with $M_{*}>10^{9.5} M_{\odot}$ is about $3500 \pm 1200 \rm kpc^{2}$,
consisting of about $55\pm20\%$ of the gas cross-section around a passive galaxy with $10^{11} M_{\odot}$. All the gas cross-section can be explained if we include the contribution of gas around satellite galaxies down to $M_{*}>10^{9} M_{\odot}$ by extrapolating the gas profiles down to the stellar mass range. 
This calculation demonstrates that cool gas of star-forming satellite galaxies is sufficient to explain a large fraction of the cool gas around passive galaxies at low redshifts. We note that this calculation does not include the contribution of stripped gas from the ISM of galaxies which is expected to further enhance the gas cross-section.
We note that \citet{Huang2016} also performed a similar calculation and argued that blue satellite galaxies alone can only account for 15\% of MgII covering fraction around luminous red galaxies, a conclusion inconsistent with our calculation. The main reason for the inconsistency between our and their results is that in \citet{Huang2016}, the MgII covering fraction is assumed to be dependent only on halo mass/stellar mass of galaxies, while in our work, we show that with the same stellar mass, the covering fraction of strong absorbers around blue galaxies is 2-4 times higher than around red galaxies at small impact parameters.

As shown in \citet{Tal2013}, the number of satellite galaxies does not evolve significantly with redshift. Therefore, to explain the redshift evolution of the power law component of the covering fraction around passive and star-forming galaxies with this satellite-origin scenario, it requires the gas distribution around each satellite galaxy becoming more extended. Such a property of gas around star-forming galaxies has been observed; the effective area of the exponential component, $f_{c}^{exp}$, becomes larger at higher redshifts and the corresponding gas cross-section, $\sigma$, scales with redshift by $\sim(1+z)^{3}$, similar to the redshift evolution of the covering fraction at large scales. This result suggests that gas associated with satellite star-forming galaxies could contribute to the gas absorption observed around both passive galaxies and star-forming galaxies and be responsible for the redshift evolution. Such a contribution of gas associated satellite galaxies to the CGM of massive galaxies is also found in galaxy simulations \citep[e.g.,][]{Hafen2019}.
However, we note that the interaction between hot and cool gas, which is expected to affect the survival of gas of satellite galaxies, is not considered here. Depending on the properties of hot gas and cool clouds, the interaction can either destroy or grow cool clouds in the halos \citep[e.g.,][]{Li2019}.

\section{Summary}
By correlating $\sim 50,000$ MgII absorbers with photometric galaxies detected in the DESI Legacy Imaging Surveys, we extract and study the properties of galaxies associated with MgII absorbers based on an effective sample of $\sim 15,000$ from redshift 0.4 to 1.3. The measurements enable us to map out the distribution of cool gas in the CGM as a function of absorption strength, galaxy types, stellar mass, impact parameter and redshift. Our findings are summarized as follows:
\begin{enumerate}

    \item The covering fraction of MgII absorbers increases with stellar mass of galaxies by $\sim M_{\odot}^{0.4}$, while after we normalize the impact parameters by the virial radius of dark matter halos, the gas distributions around galaxies with mass from $10^{9}$ to $10^{11} M_{\odot}$ become weakly dependent on stellar mass. 
    
    \item We characterize the gas profiles of strong absorbers around star-forming galaxies with a combination of an exponential and a power law profiles, one capturing the inner gas distribution and the other for the outer region. In contract,  a power law profile is sufficient to describe the gas distribution around passive galaxies. We find that within $0.3 r_{vir}$, the covering fraction of strong absorbers around star-forming galaxies is about 2-4 times higher than that around passive galaxies, demonstrating that the dichotomy of galaxy types is reflected in the cool CGM at all redshifts. 
    
    \item The covering fraction of strong absorbers ($W_{\lambda2796}>1 \rm \, \AA$) around both star-forming and passive galaxies evolves significantly with redshift, similarly to the evolution of star-formation rate of galaxies. In contrast, the covering fraction for weak absorbers ($0.4<W_{\lambda2796}<1 \rm \, \AA$) is consistent with no redshift evolution. 
    
    \item We estimate the mass of neutral hydrogen traced by strong absorbers within the virial radius of dark matter halos and show that the neutral hydrogen mass, $M_{\rm HI}$, in the CGM increases with stellar mass and evolves with redshift. The redshift evolution of $M_{\rm HI}$ is consistent with the redshift evolution of star-formation rate of galaxies. Given that the redshift evolution is observed in the CGM of both star-forming and passive galaxies, the evolution of the cool gas may reflect the global star-formation activity (i.e. including all the contribution from satellite galaxies) in the halos instead of the star-formation activity of central galaxies. 
    
    \item We estimate the neutral hydrogen mass only associated with the star-formation rate of central galaxies by subtracting the HI mass around passive galaxies from the mass around star-forming galaxies. We find that this excess mass is proportional to the SFR of galaxies, $\delta M_{\rm HI}\propto \rm SFR$. 
    This result demonstrates that the evolution of the cool gas in the CGM traces the evolution of star-formation rate of galaxies at all redshifts. In other words, galaxies and their CGM co-evolve with time. 
\end{enumerate}
These results provide novel relationships between galaxies and their CGM and quantitative constraints on models of galaxy evolution and formation, which shed new light onto the physical mechanisms that drive the cosmic baryon cycle.

The statistical technique applied in this work is powerful for extracting information from imaging surveys. It can be applied to any ongoing and future imaging surveys, such as HSC \citep{Aihara2018}, LSST \citep{Ivezic2019}, and Euclid \citep{Amiaux2012}. These deep surveys will enable us to push the study of galaxy-absorber connection to higher redshifts and fainter galaxies than the ranges probed in the work. In addition, at high redshifts, one can use different CGM tracers such as CIV \citep[
$\sim10^{4.5-5}$ K, e.g.,][]{Cooksey2013}, which is available in optical spectra at $z>1.5$. On the other hand, future large spectroscopy surveys, such as DESI \citep{Schlegel2011,Levi2013}, SDSS-V \citep{Kollmeier}, 
4MOST \citep{4most} and MOONs \citep{Moons}, will offer more spectra, increase the sample size of absorption line systems, and improve the signal-to-noise ratio of the cross-correlation measurements. 
Taken together, the combination of the technique and future datasets will enable us to probe the connection between galaxies and the multi-phase CGM and advance our understanding of galaxy formation and evolution. 


\acknowledgements

We thank Guangtun Zhu and Brice M\'enard for making their absorber catalog public available. We thank Brice M\'enard, Houjun Mo,  and J.~Xavier Prochaska for providing comments and suggestions for the early draft of this paper. 
We also thank the anonymous referee for useful comments and suggestions for improving the paper. 
We thank John Moustakas and the PRIMUS team for providing the star formation rate and stellar mass measurements of the PRIMUS sample. 
We thank the DESI Legacy Survey team for making the imaging dataset to the public. 
TWL acknowledges support from NSF grant AST-1911140.
Kavli IPMU is supported by World Premier International Research Center Initiative of the Ministry of Education, Japan.

Funding for the SDSS and SDSS-II has been provided by the Alfred P. Sloan Foundation, the Participating Institutions, the National Science Foundation, the U.S. Department of Energy, the National Aeronautics and Space Administration, the Japanese Monbukagakusho, the Max Planck Society, and the Higher Education Funding Council for England. The SDSS Web Site is http://www.sdss.org/. Funding for SDSS-III has been provided by the Alfred P. Sloan Foundation, the Participating Institutions, the National Science Foundation, and the U.S. Department of Energy Office of Science. The SDSS-III web site is http://www.sdss3.org/.
SDSS-III is managed by the Astrophysical Research Consortium for the Participating Institutions of the SDSS-III Collaboration including the University of Arizona, the Brazilian Participation Group, Brookhaven National Laboratory, University of Cambridge, Carnegie Mellon University, University of Florida, the French Participation Group, the German Participation Group, Harvard University, the Instituto de Astrofisica de Canarias, the Michigan State/Notre Dame/JINA Participation Group, Johns Hopkins University, Lawrence Berkeley National Laboratory, Max Planck Institute for Astrophysics, Max Planck Institute for Extraterrestrial Physics, New Mexico State University, New York University, Ohio State University, Pennsylvania State University, University of Portsmouth, Princeton University, the Spanish Participation Group, University of Tokyo, University of Utah, Vanderbilt University, University of Virginia, University of Washington, and Yale University.

The Legacy Surveys consist of three individual and complementary projects: the Dark Energy Camera Legacy Survey (DECaLS; NOAO Proposal ID $\#$ 2014B-0404; PIs: David Schlegel and Arjun Dey), the Beijing-Arizona Sky Survey (BASS; NOAO Proposal ID $\#$ 2015A-0801; PIs: Zhou Xu and Xiaohui Fan), and the Mayall z-band Legacy Survey (MzLS; NOAO Proposal ID $\#$ 2016A-0453; PI: Arjun Dey). DECaLS, BASS and MzLS together include data obtained, respectively, at the Blanco telescope, Cerro Tololo Inter-American Observatory, National Optical Astronomy Observatory (NOAO); the Bok telescope, Steward Observatory, University of Arizona; and the Mayall telescope, Kitt Peak National Observatory, NOAO. The Legacy Surveys project is honored to be permitted to conduct astronomical research on Iolkam Du'ag (Kitt Peak), a mountain with particular significance to the Tohono O'odham Nation.

NOAO is operated by the Association of Universities for Research in Astronomy (AURA) under a cooperative agreement with the National Science Foundation.

This project used data obtained with the Dark Energy Camera (DECam), which was constructed by the Dark Energy Survey (DES) collaboration. Funding for the DES Projects has been provided by the U.S. Department of Energy, the U.S. National Science Foundation, the Ministry of Science and Education of Spain, the Science and Technology Facilities Council of the United Kingdom, the Higher Education Funding Council for England, the National Center for Supercomputing Applications at the University of Illinois at Urbana-Champaign, the Kavli Institute of Cosmological Physics at the University of Chicago, Center for Cosmology and Astro-Particle Physics at the Ohio State University, the Mitchell Institute for Fundamental Physics and Astronomy at Texas A$\&$M University, Financiadora de Estudos e Projetos, Fundacao Carlos Chagas Filho de Amparo, Financiadora de Estudos e Projetos, Fundacao Carlos Chagas Filho de Amparo a Pesquisa do Estado do Rio de Janeiro, Conselho Nacional de Desenvolvimento Cientifico e Tecnologico and the Ministerio da Ciencia, Tecnologia e Inovacao, the Deutsche Forschungsgemeinschaft and the Collaborating Institutions in the Dark Energy Survey. The Collaborating Institutions are Argonne National Laboratory, the University of California at Santa Cruz, the University of Cambridge, Centro de Investigaciones Energeticas, Medioambientales y Tecnologicas-Madrid, the University of Chicago, University College London, the DES-Brazil Consortium, the University of Edinburgh, the Eidgenossische Technische Hochschule (ETH) Zurich, Fermi National Accelerator Laboratory, the University of Illinois at Urbana-Champaign, the Institut de Ciencies de l'Espai (IEEC/CSIC), the Institut de Fisica d'Altes Energies, Lawrence Berkeley National Laboratory, the Ludwig-Maximilians Universitat Munchen and the associated Excellence Cluster Universe, the University of Michigan, the National Optical Astronomy Observatory, the University of Nottingham, the Ohio State University, the University of Pennsylvania, the University of Portsmouth, SLAC National Accelerator Laboratory, Stanford University, the University of Sussex, and Texas A$\&$M University.

BASS is a key project of the Telescope Access Program (TAP), which has been funded by the National Astronomical Observatories of China, the Chinese Academy of Sciences (the Strategic Priority Research Program "The Emergence of Cosmological Structures" Grant $\#$ XDB09000000), and the Special Fund for Astronomy from the Ministry of Finance. The BASS is also supported by the External Cooperation Program of Chinese Academy of Sciences (Grant $\#$ 114A11KYSB20160057), and Chinese National Natural Science Foundation (Grant $\#$ 11433005).

The Legacy Survey team makes use of data products from the Near-Earth Object Wide-field Infrared Survey Explorer (NEOWISE), which is a project of the Jet Propulsion Laboratory/California Institute of Technology. NEOWISE is funded by the National Aeronautics and Space Administration.

The Legacy Surveys imaging of the DESI footprint is supported by the Director, Office of Science, Office of High Energy Physics of the U.S. Department of Energy under Contract No. DE-AC02-05CH1123, by the National Energy Research Scientific Computing Center, a DOE Office of Science User Facility under the same contract; and by the U.S. National Science Foundation, Division of Astronomical Sciences under Contract No. AST-0950945 to NOAO.

\begin{table}[ht] 
\caption{Best-fitting slopes for 50-600 kpc}
\centering

\begin{tabular}{c|c|c}
\hline  
& \multicolumn{1}{c|}{Star-forming}& \multicolumn{1}{c}{Passive galaxies}  \\
\hline
$0.4<W_{\lambda2796}<1 \rm \, \AA$ & $-1.42\pm0.06$ & $-1.01\pm0.05$ \\
$W_{\lambda2796}>1 \rm \, \AA$ & $-1.40\pm0.06$ & $-1.33\pm0.06$ \\
\hline
\end{tabular}
\label{table:slope}
\end{table}

\begin{table*}[ht] 
\caption{Best-fitting parameters for MgII covering fraction from $50<r_{p}<600$ kpc (top) and for neutral hydrogen mass within $r_{vir}$ (bottom)}
\centering

\begin{tabular}{c|ccc|ccc}
\hline  
& \multicolumn{3}{c|}{Star-forming}& \multicolumn{3}{c}{Passive galaxies}  \\
& $A$ & $\alpha$ & $\beta$ & $A$ & $\alpha$ & $\beta$  \\
\hline
$0.4<W_{\lambda2796}<1 \rm \, \AA$ & $0.026\pm0.006$ & $1.2\pm0.4$ & $0.50\pm0.05$ &
 $0.040\pm0.010$ &$-0.1\pm0.4$ & $0.34\pm0.07$ \\
$W_{\lambda2796}>1 \rm \, \AA$ & $0.008\pm0.002$ & $2.2\pm0.4$ & $0.53\pm0.05$ &  $0.007\pm0.002$ &$2.5\pm0.4$ & $0.20\pm0.07$ \\
\hline
$M_{\rm HI}(0.4<W_{\lambda2796}<1 \rm \, \AA) [10^{9} M_{\odot}]$ & $0.18\pm0.05$ & $1.6\pm0.5$ & $0.75\pm0.07$ & $0.12\pm0.04$ &$0.3\pm0.5$ & $1.17\pm0.12$ \\
$M_{\rm HI}(W_{\lambda2796}>1 \rm \, \AA) [10^{9} M_{\odot}]$ & $0.23\pm0.06$ & $3.1\pm0.4$ & $0.64\pm0.06$ & $0.10\pm0.03$ &$2.9\pm0.5$ & $0.95\pm0.12$ \\
\hline
\end{tabular}
\label{table:a_100kpc}
\end{table*}

\begin{table*}[ht] 
\caption{Best-fitting parameters for MgII covering fraction normalized by virial radius}
\centering

\begin{tabular}{cc|cccccc}
\hline  

& & $A$ & $\alpha$ & $\beta$ & $\gamma$ & $c$ & $d$ \\
\hline
\multirow{2}{*}{Star-forming}&$0.4<W_{\lambda2796}<1 \rm \, \AA$ & $0.014\pm0.003$ & $1.4\pm0.4$ & $0.18\pm0.04$ & $-1.14\pm0.05$ & - & - \\
    &$W_{\lambda2796}>1 \rm \, \AA$ & $0.0024\pm0.0011$ & $3.0\pm0.8$ & $0.11\pm0.07$ & $-0.94\pm0.20$ & $0.06\pm0.01$ & $1.4\pm0.3$ \\
\hline    
\multirow{2}{*}{Passive}&$0.4<W_{\lambda2796}<1 \rm \, \AA$ & 
$0.026\pm0.006$ & $0.8\pm0.4$ & $-0.32\pm0.06$ & $-0.77\pm0.05$ & -  & - \\
    &$W_{\lambda2796}>1 \rm \, \AA$  & $0.004\pm0.001$ & $4.0\pm0.4$ & $-0.64\pm0.06$ & $-1.08\pm0.05$ &-& -\\


\hline
\end{tabular}
\label{table:a_rv}
\end{table*}

\appendix
\section{A correction function for galaxy number counts at small scales}
\begin{figure*}
\center
\includegraphics[width=1\textwidth]{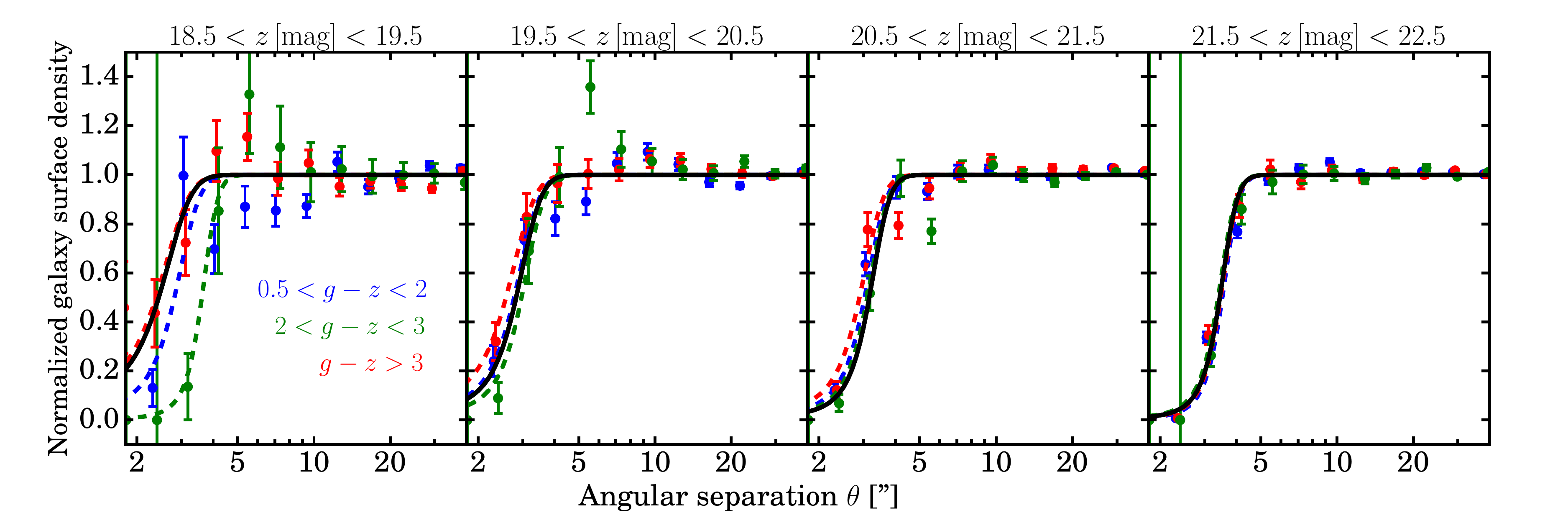}
\caption{Normalized surface density of galaxies around quasars at $z>1.6$. The brightness of galaxies, listed on the top of each panel, increases from left panel to right panel. The colors indicate three galaxy color bins, $0.5<g-z<2$ (blue), $2<g-z<3$ (green), and $g-z>3$ (red). The color dashed lines show the individual best-fit functions and the black solid lines illustrate the global best-fit function.}
\label{fig:correction}
\end{figure*}

When two objects are close to each other on the sky, they will blend together and the imaging algorithm could fail to de-blend the two objects. In this analysis, faint galaxies close to bright quasars may not be detected by the algorithm due to the blending effect and therefore directly counting number of detected galaxies at small angular scales will underestimate the intrinsic galaxy number. To correct for this blending effect, we empirically estimate the probability of detecting galaxies near bright quasars. We use our reference quasars with targeted absorbers at redshifts higher than 1.6. The selection ensures that all the quasars are at redshift higher than 1.6 so that galaxies physically clustering with quasars do not affect the measurements. 
We search photometric galaxies in the Legacy Surveys around the quasar sample, and calculate the surface number density of galaxies as a function of angular separation, galaxy brightness, and color. Figure~\ref{fig:correction} show the surface number density of galaxies as a function of angular separation from the central quasars. The surface number density of galaxies is normalized by the surface number density obtained at $75"<\theta<100"$.

As can be seen in Figure~\ref{fig:correction}, the surface number density of galaxies is much lower than 1 within $3"$ of quasars due to the blending effect. 
We find that the normalized galaxy number can be described by $1/w$ where $w$ is the correction function. The function can be parameterized as
\begin{equation}
    w(\theta) = 1+C\times e^{-\theta^{2}/2}.
    \label{eq:correction1}
\end{equation}
The color dashed lines in Figure~\ref{fig:correction} show the best-fit for each galaxy color bin and magnitude bin. We find that the galaxy detection rate weakly depends on galaxy color but depends on galaxy brightness. Therefore, we obtain a global best-fit correction function,
\begin{equation}
    w({\rm mag},\theta) = 1+(54\pm5)\times\bigg(\frac{\rm mag}{20}\bigg)^{20.3\pm1.3}\times e^{-\theta^{2}/2},
    \label{eq:correction}
\end{equation}
which is shown by the black lines. In this work, we use Eq.~\ref{eq:correction} to statistically recover the galaxy number counts close to quasars. 

\section{Covering fraction as a function of stellar mass and redshift}
\begin{figure*}
\center
\includegraphics[width=1\textwidth]{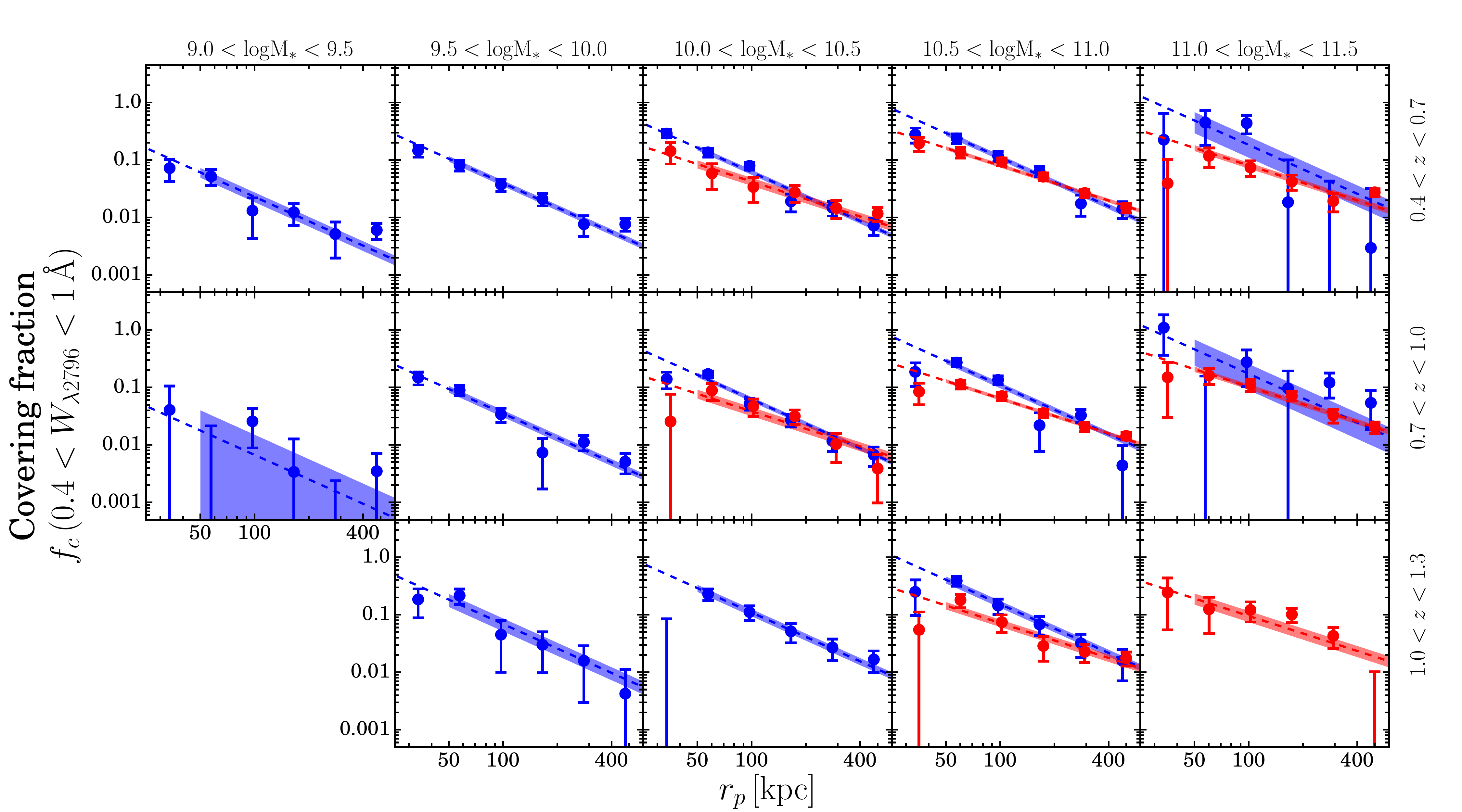}
\includegraphics[width=1\textwidth]{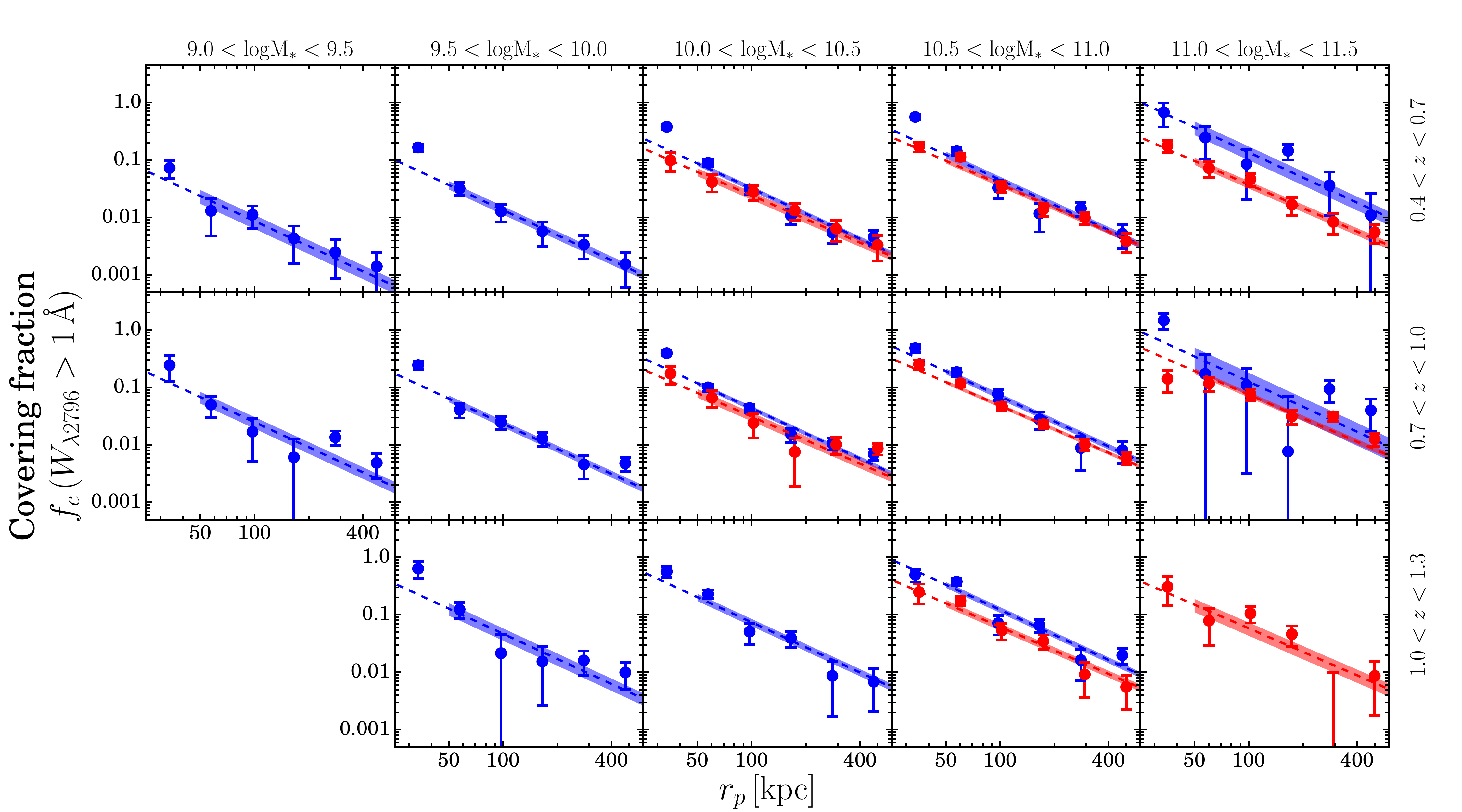}
\caption{MgII covering fraction of weak absorbers ($0.4<W_{\lambda2796}<1 \rm \, \AA$) (upper panel) and strong absorbers ($W_{\lambda2796}>1 \rm \, \AA$) (lower panel) from 20 kpc to 600 kpc as a function of galaxy stellar mass and redshift. The range of stellar mass is listed on the top and increases from left to right. The redshift range is listed on the right and increases from top to bottom. Star-forming galaxies and passive galaxies are indicated by the blue and red data points respectively. The best-fit power laws are indicated by the color bands.}
\label{fig:fc}
\end{figure*}

Here we present the measured covering fraction as a function of stellar mass and redshift. In this paper, we focus on stellar mass bins having the signal-to-noise ratios of the mean number of galaxies associated with absorbers, $\langle N_{\rm gal}^{\rm abs}\rangle$, measured within 200 kpc greater than 2. Figure~\ref{fig:fc} shows the covering fraction of weak absorbers ($0.4<W_{\lambda2796}<1 \rm \, \AA$, upper panel) and strong absorbers ($W_{\lambda2796}>1 \rm \, \AA$, lower panel) around star-forming (blue data points) and passive galaxies (red data points) from 20 kpc up to 600 kpc. The stellar mass of galaxies increases from left to right and the redshift from top to bottom. We fit each covering fraction with a power law profile (Equation 6) and the best-fitting values are listed in Table~\ref{table:slope} and Table~\ref{table:a_100kpc}.
The best-fit power laws are shown in Figure~\ref{fig:fc} with the color bands indicating $1\sigma$ uncertainties. We exclude the covering fraction at impact parameters $r_{p}<50$ kpc while obtaining the best-fits given that the excess absorption in the inner region of the cool CGM of star-forming galaxies tends to depart from the gas absorption at larger scales. Such a behavior can be observed in Figure~\ref{fig:fc} and previous studies \citep[e.g., see Fig 8 in][]{Lan2014}. We note that all the best-fit parameter values listed in the paper are obtained from measurements with finer stellar mass and redshift bins, with 0.3 dex interval for stellar mass and 0.15 interval for redshift, than the binning showed in the figures which is chosen for clear presentation.

{}

\end{CJK*}
\end{document}